\documentclass[11pt]{JHEP3}

\usepackage{amsmath}
\usepackage{amsfonts}
\usepackage{mathrsfs}
\usepackage{array}
\usepackage{epsfig}

\newcommand\MM{\mathcal{M}}

\newcommand\EE{\mathcal{E}}

\newcommand\dM{\partial \MM}

\newcommand\nts{\negthickspace}
\newcommand\bns{\nts \nts \nts}

\newcommand\al{\alpha}

\newcommand\ga{\gamma}
\newcommand\de{\delta}

\DeclareMathAlphabet{\mathpzc}{OT1}{pzc}{m}{it}

\newcommand{\eq}[2]{\begin{equation} #1 \label{#2} \end{equation}}

\newcommand{\blist}{\begin{itemize}}
\newcommand{\elist}{\end{itemize}}

\newcommand{\Rthree}{{\cal R}}
\newcommand{\Kthree}{{\cal K}}

\newcommand{\formerg}{g_+}

\preprint{MIT-CTP 3981}

\title{Holographic Description of AdS$_2$ Black Holes}
\author{Alejandra Castro \\
          Department of Physics, University of Michigan,\\ 431 Church Street,
Ann Arbor, MI 48109, USA.\\
          Email: \email{aycastro@umich.edu}}

\author{Daniel Grumiller\\
          Center for Theoretical Physics,
          Massachusetts Institute of Technology,\\
          77 Massachusetts Ave.,
          Cambridge, MA  02139, USA \\
          and \\
          Institute for Theoretical Physics,
      Vienna University of Technology,\\
          Wiedner Hauptstr. 8-10/136, A-1040 Vienna, Austria\\
          Email: \email{grumil@lns.mit.edu}}

\author{Finn Larsen\\
          Department of Physics, University of Michigan,\\  431 Church Street,
Ann Arbor, MI 48109, USA.\\
and \\Theory Division, CERN, CH-1211 Geneva 23, Switzerland.\\
          Email: \email{larsenf@umich.edu}}

\author{Robert McNees\\
         Perimeter Institute for Theoretical Physics, \\
         31 Caroline Street North,
         Waterloo, Ontario N2L 2Y5, Canada \\
         Email: \email{mcnees@perimeterinstitute.ca}}

\keywords{holographic renormalization, boundary counterterms, two-dimensional Maxwell-dilaton gravity, AdS black hole thermodynamics}

\abstract{
We develop the holographic renormalization of AdS$_2$ gravity systematically. We
find that a bulk Maxwell term necessitates a boundary mass term for the gauge field
and verify that this unusual term is invariant under gauge transformations that preserve
the boundary conditions. We determine the energy-momentum tensor and the central
charge, recovering recent results by Hartman and Strominger. We show that our
expressions are consistent with dimensional reduction of the AdS$_3$ energy-momentum
tensor and the Brown--Henneaux central charge. As an application of our results
we interpret the entropy of AdS$_2$ black holes as the ground state entropy of a dual CFT.
}

\begin{document}

\section{Introduction}
\label{sec:Intro}

Extremal black hole spacetimes universally include an AdS$_2$ factor
\cite{Kunduri:2007vf}. It is therefore natural to study quantum black
holes by applying the AdS/CFT correspondence to the AdS$_2$ factor.
There have been several interesting attempts at implementing this
strategy \cite{Strominger:1998yg,Maldacena:1998uz,Michelson:1999dx,
Brigante:2002rv,Astorino:2002bj,Verlinde:2004gt,Gupta:2008ki,Alishahiha:2008tv,Sen:2008vm}
but AdS$_2$ holography remains enigmatic, at least compared with the
much more straightforward case of AdS$_3$ holography.

Recently a new approach was proposed by Hartman and Strominger
\cite{Hartman:2008dq}, in the context of Maxwell-dilaton gravity with
bulk action \eq{ I_{\rm bulk} = \frac{\alpha}{2\pi} \int_{\MM}\!\!\!
d^2x\sqrt{-g}\, \left[
e^{-2\phi}\left(R+\frac{8}{L^2}\right)-\frac{L^2}{4} \,F^2 \right] ~.
}{eq:aa} These authors pointed out that, for this theory, the usual
conformal diffeomorphisms must be accompanied by gauge
transformations, in order to maintain boundary conditions. They found
that the combined transformations satisfy a Virasoro algebra with a
specific central charge. These results suggest a close relation to
the AdS$_3$ theory.

In this paper we develop the holographic description of AdS$_2$ for
the theory \eqref{eq:aa} systematically, following the procedures
that are well-known from the AdS/CFT correspondence in higher
dimensions. Specifically, we consider:
\begin{enumerate}
\item
{\bf Holographic renormalization.} We apply the standard holographic renormalization
procedure \cite{Balasubramanian:1999re,Emparan:1999pm,Myers:1999psa,deHaro:2000xn} to asymptotically AdS$_2$ spacetimes. In particular, we impose precise boundary
conditions and determine the boundary counterterms needed for a consistent variational
principle. These counterterms encode the infrared divergences of the bulk theory.
\item
{\bf Stress tensor and central charge.} The asymptotic
$SL(2,\mathbb{R})$ symmetry of the theory is enhanced to a Virasoro
algebra, when the accompanying gauge transformation is taken into
account. We determine the associated boundary stress tensor and its
central charge. Our result for the central charge
\eq{
c=\frac{3}{2}\,kE^2L^4
}{eq:c}
is consistent with that of Hartman and Strominger \cite{Hartman:2008dq}.
\item {\bf Dimensional reduction from 3D to 2D.} We show that our
    results in two dimensions (2D) are consistent with
    dimensional reduction of standard results in three dimensions
    (3D). In particular, we verify that our result \eqref{eq:c}
    agrees with the Brown--Henneaux central charge for AdS$_3$
    spacetimes \cite{Brown:1986nw}.
\item {\bf Entropy of AdS$_2$ black holes}. We use our results to
    discuss the entropy of black holes in AdS$_2$. To be more
    precise, we use general principles to determine enough
    features of the microscopic theory that we can determine its
    entropy, but we do not discuss detailed implementations in
    string theory. This is in the spirit of the well-known
    microscopic derivation of the entropy of the BTZ black hole
    in 3D \cite{Strominger:1997eq}, and also previous related
    results in AdS$_2$
    \cite{Maldacena:1998uz,Cadoni:1998sg,Cadoni:1999ja,NavarroSalas:1999up}.
\end{enumerate}
The main lesson we draw from our results is that, even for AdS$_2$,
the AdS/CFT correspondence can be implemented in a rather
conventional manner.

In the course of our study we encounter several subtleties. First of
all, we find that the coupling constant $\alpha$ in \eqref{eq:aa}
must be negative. We reach this result by imposing physical
conditions, such as positive central charge, positive energy,
sensible thermodynamics, and a consistent 3D/2D reduction. The
redundancy gives us confidence that we employ the physically correct
sign.

Another surprise is that consistency of the theory requires the
boundary term
\eq{ I_{\rm new} \sim \int_{\dM}\bns d x\sqrt{-\ga} \,m A^a A_a\,,
}{eq:ab}
where $m$ is a constant that we compute. The boundary term
\eqref{eq:ab} takes the form of a mass term for the gauge field. This
is remarkable, because it appears to violate gauge invariance.
However, we demonstrate that the new counterterm \eqref{eq:ab} is
invariant with respect to all gauge variations that preserve the
boundary conditions.

Some other important issues relate to the details of the KK-reduction. In our embedding
of asymptotically AdS$_2$ into AdS$_3$ we maintain Lorentzian signature and reduce
along a direction that is light-like in the boundary theory, but space-like in the bulk. 
A satisfying feature of the set-up is that the null reduction on the
boundary manifestly freezes the holomorphic sector of the boundary
theory in its ground state, as it must since the global symmetry is
reduced from $SL(2,\mathbb{R})\times SL(2,\mathbb{R})$ to
$SL(2,\mathbb{R})$. The corollary is that the boundary theory dual to
asymptotically AdS$_2$ necessarily becomes the chiral part of a CFT
and such a theory is not generally consistent by itself
\cite{Witten:2007kt,Li:2008dq}. The study of the ensuing microscopic
questions is beyond the scope of this paper.

This paper is organized as follows. In section \S{\ref{sec:2}} we set
up our model, the boundary conditions, and the variational principle.
We use this to determine the boundary counterterms and verify gauge
invariance of the mass term \eqref{eq:ab}. In section
\S{\ref{sec:BSTcc}}, we use the full action, including counterterms,
to derive the renormalized energy-momentum tensor, and the central
charge. We compare to the result of Hartman and Strominger, being
careful to spell out conventions. In section \S{\ref{sec:5}} we
present the reduction from 3D to 2D, give the identification between
fields, and verify consistency with standard results in AdS$_3$
gravity. In section \S{\ref{sec:6}} we apply our results to black
hole thermodynamics. This provides the setting for our discussion of
black hole entropy. Section \S{\ref{sec:new}} generalizes our results
to linear dilaton backgrounds and shows consistency with the constant
dilaton sector. In section \S{\ref{sec:7}} we discuss a few
directions for future research. Our conventions and notations are
summarized in appendix \ref{app:not}, and some calculations
concerning the dictionary between 3D and 2D are contained in appendix
\ref{app:A}.


\section{Boundary Counterterms in Maxwell-Dilaton AdS Gravity}
\label{sec:2}

In this section we study a charged version of a specific 2D dilaton
gravity. We construct a well-defined variational principle for this
model by adding boundary terms to the standard action, including a
novel boundary mass term for the $U(1)$ gauge field.

\subsection{Bulk action and equations of motion}\label{sec:2.1}

There exist many 2D dilaton gravity models that admit an AdS ground
state (see \cite{Grumiller:2002nm,Grumiller:2006rc} and references
therein). For the sake of specificity we pick a simple example ---
the Jackiw--Teitelboim model \cite{JT} --- and add a minimally
coupled $U(1)$ gauge field. The bulk action
\begin{equation}\label{eq:bulkAction}
I_{\rm bulk} = \frac{\alpha}{2\pi}\,\int_{\MM} \nts \nts d^{\,2}x
\sqrt{-g}\,\left[ e^{-2\phi}\,\left(R + \frac{8}{L^2}\right) -
\frac{L^{2}}{4}\,F^2 \right]~,
\end{equation}
is normalized by the dimensionless constant $\alpha$ which is left
unspecified for the time being. For constant dilaton backgrounds we
eventually employ the relation
\begin{equation}\label{eq:G2}
\alpha=-{1\over8G_2}e^{2\phi}
\end{equation}
between the 2D Newton constant $G_2$ and $\alpha$. While the factors
in \eqref{eq:G2} are the usual ones (see appendix \ref{app:not}), the
sign will be justified in later sections by computing various
physical quantities.

The variation of the action with respect to the fields takes the form
\begin{equation}\label{Variation}
\delta \,I_{\rm bulk} = \frac{\alpha}{2\pi\,}\,\int_{\MM} \nts\nts
d^{\,2}x \sqrt{-g}\,\Big[\EE^{\mu\nu}\,\delta g_{\mu\nu} +
\EE_{\phi} \, \delta \phi + \EE^{\mu}\,\delta A_{\mu} \Big] + {\rm
boundary\,\,terms}~,
\end{equation}
with
\begin{subequations}
\label{eq:eom}
\begin{align}
\EE_{\mu\nu}  &= \nabla_{\mu} \nabla_{\nu} e^{-2 \phi} - g_{\mu\nu} \, \nabla^{2} e^{-2 \phi}
+ \frac{4}{L^2}\,e^{-2 \phi}\,g_{\mu\nu} + \frac{L^2}{2}\,F_{\mu}{}^\lambda\,F_{\nu\lambda} -
\frac{L^2}{8}\,g_{\mu\nu} \, F^{2}~, \label{eq:eom1} \\
\EE_{\phi} &= -2\,e^{-2 \phi}\,\left( R + \frac{8}{L^2}\right)~, \label{eq:eom2} \\
\EE_{\mu} &= L^2\,\nabla^{\nu} F_{\nu\mu} ~. \label{eq:eom3}
\end{align}
\end{subequations}
Setting each of these equal to zero yields the equations of motion
for the theory. The boundary terms will be discussed in
section \S\ref{sec:2.2} below.

All classical solutions to \eqref{eq:eom} can be found in closed form
\cite{Grumiller:2002nm,Grumiller:2006rc}. Some aspects of generic
solutions with non-constant dilaton will be discussed in section
\S\ref{sec:new}, below. Until then we focus on solutions with
constant dilaton, since those exhibit an interesting enhanced
symmetry. This can be seen by noting that the dilaton equation
$\EE_\phi=0$ implies that all classical solutions must be spacetimes
of constant (negative) curvature. Such a space is maximally symmetric
and exhibits three Killing vectors, {\it i.e.}~it is locally (and
asymptotically) AdS$_2$. A non-constant dilaton breaks the
$SL(2,\mathbb{R})$ algebra generated by these Killing vectors to
$U(1)$, but a constant dilaton respects the full AdS$_2$ algebra.

With constant dilaton the equations of motion reduce to
\begin{gather}\label{redEqns}
 R + \frac{8}{L^2} = 0~, \quad \quad \nabla^{\nu} F_{\nu\mu} = 0~, \quad \quad e^{-2 \phi} = -\frac{L^4}{32}\,F^2  ~.
\end{gather}
The middle equation in \eqref{redEqns} is satisfied by a covariantly
constant field strength
\begin{equation}
F_{\mu\nu} =  2E\, \epsilon_{\mu\nu},
\label{eq:dg1}
\end{equation}
where $E$ is a constant of motion determining the strength of the electric field.
The last equation in \eqref{redEqns} determines the dilaton in terms
of the electric field,
\begin{equation}
e^{-2\phi} =  \frac{L^4}{4}\,E^2\,.
\label{eq:dilaton}
\end{equation}
Expressing the electric field in terms of the dilaton, we can
rewrite \eqref{eq:dg1} as $F_{\mu\nu}=\frac{4}{L^2}\,
e^{-\phi}\,\epsilon_{\mu\nu}$. Without loss of generality, we have
chosen the sign of $E$ to be positive. The first equation in
\eqref{redEqns} requires the scalar curvature to be constant and
negative. Working in a coordinate and $U(1)$ gauge where the metric
and gauge field take the form
\begin{equation}
 ds^{2} =  d\eta^{2} + g_{tt}\,dt^{2} = d\eta^{2} + h_{tt}\,dt^{2}~, \quad \quad A_{\mu} dx^{\mu} = A_{t}(\eta,t)\,dt
 ~,
 \label{gauge}
\end{equation}
the curvature condition simplifies to the linear differential
equation
\begin{equation}
\frac{\partial^2 }{\partial \eta^2}\sqrt{-g} =
\frac{4}{L^2}\,\sqrt{-g}\,, \label{eq:dg2}
\end{equation}
which is solved by $\sqrt{-g}=\big(h_0(t)\,e^{2\eta/ L} +
h_1(t)\,e^{-2\eta/L}\big)/2$. Therefore, a general solution to
\eqref{redEqns} is given by
\begin{subequations}\label{Solution}
\begin{align}\label{metricsolution}
g_{\mu\nu} dx^{\mu} dx^{\nu} &= d\eta^{2} -  \frac{1}{4}\,\left(h_0(t)\,e^{2\eta/ L}
+ h_1(t)\,e^{-2\eta/L} \right)^{2}\,dt^2~, \\ \label{MaxwellSolution}
A_{\mu} dx^{\mu} & =
\frac{1}{L}\,e^{-\phi}\,\left(h_0(t)\,e^{2\eta/L} - h_1(t)\,e^{-2\eta/L} +
a(t) \right)\,dt~, \quad \quad \\
\phi &= \rm constant~,
\end{align}
\end{subequations}
where $h_0$, $h_1$, and $a$ are arbitrary functions of $t$. This
solution can be further simplified by fixing the residual gauge
freedom in \eqref{gauge}. In particular, the $U(1)$ transformation
$A_{\mu} \to A_{\mu} + \partial_{\mu} \Lambda(t)$ preserves the
condition $A_{\eta} = 0$, and a redefinition $h_0(t)dt \to dt$ of
the time coordinate preserves the conditions $g_{\eta\eta} = 1$ and
$g_{\eta t} = 0$. This remaining freedom is fixed by requiring
$a(t)=0$ and $h_0(t)=1$. Thus, the general gauge-fixed solution of
the equations of motion depends on the constant $\phi$, specified by
the boundary conditions, and an arbitrary function $h_1(t)$.

Following the standard implementation of the AdS/CFT correspondence
in higher dimensions, we describe asymptotically AdS$_2$ field
configurations by \eqref{gauge} with the Fefferman-Graham expansions:
\begin{subequations}\label{eq:FG}
\begin{align}
h_{tt} &=  e^{4\eta/L}\,g_{tt}^{(0)} + g_{tt}^{(1)} + e^{-4\eta/L}\,g_{tt}^{(2)}+\ldots~,  \label{eq:FGmetric} \\
A_t & = e^{2\eta/L}\,A_t^{(0)} + A_t^{(1)}+e^{-2\eta/L}\, A_t^{(2)}+\ldots~, \label{eq:FGgf} \\
\phi &= \phi^{(0)} + e^{-2\eta/L}\,\phi^{(1)}+\ldots
\label{eq:FGdil} ~.
\end{align}
\end{subequations}
Our explicit solutions \eqref{Solution} take this form with asymptotic values
\begin{equation}
g_{tt}^{(0)} = -\frac14 \,,\qquad A_t^{(0)} = \frac{1}{L}\,e^{-\phi^{(0)}}\,,\qquad \phi^{(0)} = \rm constant\,,
\label{eq:bc}
\end{equation}
and specific values for the remaining expansion coefficients in \eqref{eq:FG}. The variational
principle considers general off-shell field configurations with \eqref{eq:bc} imposed
as boundary conditions, but the remaining expansion coefficients are free to vary from
their on-shell values.

\subsection{Boundary terms}\label{sec:2.2}

An action principle based on \eqref{eq:bulkAction} requires a number of boundary terms:
\begin{equation}
I=I_{\rm bulk} + I_{\rm GHY} + I_{\rm counter} = I_{\rm bulk} + I_{\rm boundary}~.
\label{eq:bulkandboundary}
\end{equation}
The boundary action $I_{\rm GHY}$ is the dilaton gravity analog of
the Gibbons--Hawking--York (GHY) term
\cite{Gibbons:1976ue,York:1972sj}, and it is given by
\begin{equation}
I_{\rm GHY} = \frac{\alpha}{\pi}\,\int_{\dM} \bns dx\sqrt{-h}\,e^{-2\phi}\,K\,,
\label{eq:GHY}
\end{equation}
where $h$ is the determinant of the induced metric on $\dM$, and $K$
the trace of the extrinsic curvature (our conventions are summarized
in appendix \ref{app:not}). This term is necessary for the action to
have a well-defined boundary value problem for fields satisfying
Dirichlet conditions at $\dM$. However, on spacetimes with
non-compact spatial sections this is not sufficient for a consistent
variational principle. We must include in \eqref{eq:bulkandboundary}
a set of `boundary counterterms' so that the action is extremized by
asymptotically AdS$_2$ solutions of the equations of motion. In order
to preserve the boundary value problem these counterterms can only
depend on quantities intrinsic to the boundary. Requiring
diffeomorphism invariance along the boundary leads to the generic
{\it ansatz}
\begin{equation}
I_{\rm counter} = \int_{\dM}\bns dx \sqrt{-h}\,{\cal L}_{\rm counter}(A^a A_a,\phi)\,.
\label{eq:counterformal}
\end{equation}
In the special case of vanishing gauge field the counterterm must
reduce to ${\cal L}_{\rm counter}\propto e^{-2\phi}$,
cf.~e.g.~\cite{Grumiller:2007ju}. In the presence of a gauge field
the bulk action contains a term that scales quadratically with the
field strength. Therefore, the counterterm may contain an additional
contribution that scales quadratically with the gauge field. This
lets us refine the {\it ansatz} \eqref{eq:counterformal} to
\begin{equation}
I_{\rm counter} = \frac{\alpha}{\pi}\,\int_{\dM}\bns dx
\sqrt{-h}\,\Big[\lambda\,e^{-2\phi} + m\,A^a A_a\Big]\,.
\label{eq:counter}
\end{equation}
The coefficients $\lambda, m$ of the boundary counterterms will be
determined in the following.

With these preliminaries the variation of the action
\eqref{eq:bulkandboundary} takes the form
\begin{equation}
\delta \,I = \int_{\dM} \bns dx \sqrt{-h}\,\Big[(\pi^{ab} +
p^{ab})\,\delta h_{ab} + (\pi_{\phi} + p_{\phi})\,\delta \phi +
(\pi^{a} + p^{a})\,\delta A_a \Big] + {\rm bulk\,\,terms}\,,
\label{eq:variation}
\end{equation}
where the bulk terms were considered already in the variation of the
bulk action \eqref{Variation}. The boundary contributions are given
by
\begin{subequations}\label{eq:bndymom}
\begin{align}\label{hmomentum}
\pi^{tt} + p^{tt} & \, = \frac{\alpha}{2\pi}\,\big(
h^{tt}\,n^{\mu}\partial_{\mu} e^{-2\phi} + \lambda \, h^{tt}\,e^{-2
\phi} + m\,h^{tt}\,A^{t} A_t - 2\,m\,A^{t} A^{t}\big)~, \\
\label{Amomentum} \pi^{t} + p^{t} & \, = \frac{\alpha}{2\pi}\,\big(
-L^{2}\,n_{\mu}\,F^{\mu t} + 4\,m\,A^{t}\big)~, \\
\label{phimomentum} \pi_{\phi} + p_{\phi} & = -2
\frac{\alpha}{\pi}\,e^{-2 \phi}\,\big( K + \lambda \big) ~.
\end{align}
\end{subequations}
In our notation `$\pi$' is the part of the momentum that comes from
the variation of the bulk action and the GHY term, and `$p$'
represents the contribution from the boundary counterterms.

For the action to be extremized the terms in \eqref{eq:variation}
must vanish for generic variations of the fields that preserve the
boundary conditions \eqref{eq:bc}. If we consider field
configurations admitting an asymptotic expansion of the form
\eqref{eq:FG}, then the boundary terms should vanish for arbitrary
variations of the fields whose leading asymptotic behavior is:
\begin{subequations}
\label{eq:boundaryvariations}
\begin{align}
\delta h_{tt} & = \delta g_{tt}^{(1)} = \rm finite \\
\delta A_t & = \delta A_t^{(1)} = \rm finite \label{eq:gaugebcond} \\
\delta \phi & = e^{- 2\eta/L}\,\delta \phi^{(1)} \to 0
\end{align}
\end{subequations}
We refer to variations of the form \eqref{eq:boundaryvariations}
as ``variations that preserve the boundary conditions''.

Inserting the asymptotic behavior \eqref{eq:FG} in
\eqref{hmomentum}-\eqref{phimomentum}, the boundary terms in
\eqref{eq:variation} become
\begin{align}
\left. \delta I \raisebox{11pt}{\,} \right|_\textrm{\tiny EOM} =
{\alpha\over\pi}\int_{\dM} \bns dt \, & \, \Big[
-\,e^{-2\phi}\,\left(\lambda+{4\over L^2} m\right)\,e^{- 2
\eta/L}\,\delta h_{tt}
- e^{-2\phi}\,\left({2\over L}+\lambda\right)\,e^{2\eta/L}\,\delta \phi \nonumber \\
 & \quad +2e^{-\phi}\,\left(1-{2\over L} m\right)\, \delta A_t + \ldots  \Big]~,
 \label{bndyvar}
\end{align}
where `$\ldots$' indicates terms that vanish at spatial infinity for
any field variations that preserve the boundary conditions. The
leading terms in \eqref{bndyvar} vanish for $\lambda$ and $m$
given by
\begin{gather}\label{counterterms}
 \lambda = - \frac{2}{L}~, \quad \quad \quad m = \frac{L}{2} ~.
\end{gather}
As a consistency check we note that these {\em two} values cancel
{\em three} terms in \eqref{bndyvar}. Also, the value of $\lambda$,
which is present for dilaton gravity with no Maxwell term, agrees
with previous computations \cite{Grumiller:2007ju}. \footnote{We also
comment on the only previous example of $A^2$ boundary terms that we
are aware of \cite{Martelli:2002sp}. That work employs the Einstein
frame, which is not accessible in 2D, and many of the expressions
appearing in that paper indeed diverge when applied to 2D. An
exception is their equation (92), which determines the numerical
factor $N_0$ in the boundary mass term (90) for the gauge field
$B_i$. Equation (92) has two solutions, and the authors of
\cite{Martelli:2002sp} exclusively consider the trivial one $N_0=0$,
{\it i.e.} there is no boundary mass term. However, the other
solution leads to a non-vanishing boundary mass term for the gauge
field. Translating their notations to ours ($d=1$, $N_0=2m
\alpha/\pi$, $K_0=\alpha L^2/(2\pi)$, $\ell=L/2$) we find perfect
agreement between the non-trivial solution $N_0=K_0/\ell$ of their
equation (90) and our result \eqref{counterterms}.} With the values
\eqref{counterterms}  the variational principle is well-defined
because the variation of the on-shell action
vanishes for all variations that preserve the boundary conditions.

In summary, the full action
\begin{align}
I= \frac{\alpha}{2\pi}\,\int_{\MM} \nts \nts d^{\,2}x \sqrt{-g} \,&
\left[ e^{-2\phi}\,\left(R + \frac{8}{L^2}\right) -
\frac{L^{2}}{4}\,F^{\mu\nu} F_{\mu\nu} \right]
\nonumber \\
+ \frac{\alpha}{\pi}\,\int_{\dM} \bns dx\,\sqrt{-h} \,&
\left[e^{-2\phi}\,\left(K - \frac{2}{L}\right) +\frac L2 A^a A_a
\right]\,, \label{Action}
\end{align}
has a well-defined boundary value problem, a
well-defined variational principle, and is extremized by
asymptotically AdS$_2$ solutions of the form \eqref{eq:FG}.

\subsection{Boundary mass term and gauge invariance}
\label{sec:gaugeinv}

The boundary term
\eq{ I_{\rm new} = \frac{\alpha L}{2\pi} \int_{\dM}\bns
dx\sqrt{-h}\,A^a A_a\ }{eq:bm25}
is novel and requires some attention, because it would seem to spoil
invariance under the gauge transformations
\eq{ A_\mu \to A_\mu +
\partial_\mu \Lambda~. }{eq:bm26}
The purpose of this section is to show that the mass term
\eqref{eq:bm25} is in fact invariant under gauge transformations that
preserve the gauge condition $A_\eta=0$ and the boundary condition
specified in \eqref{eq:gaugebcond}.

The gauge parameter $\Lambda$ must have the
asymptotic form
 \eq{
\Lambda = \Lambda^{(0)}(t) + \Lambda^{(1)}(t)\,e^{-2\eta/L} + {\cal
O}\left(e^{-4\eta/L}\right)}{eq:bm27}
in order that the asymptotic behavior
\begin{subequations}
\label{eq:gt}
\begin{align}
A_t &= A_t^{(0)}\,e^{2\eta/L} + {\cal O}(1) ~,
\label{eq:gt1}
 \\
A_\eta &= {\cal O}\left(e^{-2\eta/L}\right) ~,
\label{eq:gt2}
\end{align}
\end{subequations}
of the gauge field is preserved. Indeed, allowing some positive power of
$e^{2\eta/L}$ in the expansion \eqref{eq:bm27}  of
$\Lambda$ would spoil this property.

Having established the most general gauge transformation consistent
with our boundary conditions we can investigate whether the
counterterm \eqref{eq:bm25} is gauge invariant. Acting with the gauge
transformation \eqref{eq:bm26} and taking the asymptotic expansions
\eqref{eq:bm27} and \eqref{eq:gt} into account yields
\eq{ \de_\Lambda I_{\rm new} =  \frac{\alpha L}{\pi}
\lim_{\eta\to\infty} \int_{\dM}\bns dt \sqrt{-h}\, h^{tt} A_t
\,\de_\Lambda A_t = -\frac{2\alpha L}{\pi}\,A_t^{(0)}\,
\int_{\dM}\bns dt \,\partial_t \Lambda^{(0)}~. }{eq:bm28} The same
result holds for the full action \eqref{Action}, because all other
terms in $I$ are manifestly gauge invariant. The integral in
\eqref{eq:bm28} vanishes for continuous gauge transformations if
$\Lambda^{(0)}$ takes the same
value at the initial and final times. In those cases the counterterm
\eqref{eq:bm25} and the full action \eqref{Action} are both gauge
invariant with respect to gauge transformations that asymptote to
\eqref{eq:bm27}.

The ``large'' gauge transformations that do not automatically leave
the action invariant are also interesting. As an example, we consider
the discontinuous gauge transformation
\eq{ \Lambda^{(0)}(t) = 2\pi\,q_m \,\theta(t-t_0)~, }{eq:bm29}
where $q_m$ is the dimensionless magnetic monopole charge with a
convenient normalization. We assume that $t_0$ is contained in $\dM$,
so that the delta function obtained from $\partial_t\Lambda^{(0)}$ is
supported. Inserting the discontinuous gauge transformation
\eqref{eq:bm29} into the gauge variation of the action
\eqref{eq:bm28} gives
\eq{
\de_\Lambda I = \de_\Lambda I_{\rm new} = -2\al\,L^2E\,q_m~,
}{eq:bm30}
which tells us that the full action is shifted by a constant. We
investigate now under which conditions this constant is an integer
multiple of $2\pi$.

The 2D Gauss law relates the electric field $E$ to the dimensionless electric charge $q_e$:
\begin{equation}
E = -\frac{\pi\,q_e}{\al\, L^2}~.
\label{eq:Gauss}
\end{equation}
Again we have chosen a convenient normalization.\footnote{If we set
$\alpha\,L^2/(2\pi)=1$ then the action \eqref{eq:aa} has a
Maxwell-term with standard normalization. In that case our Gauss law
\eqref{eq:Gauss} simplifies to $2E=-q_e$. The factor 2 appears here
because in our conventions the relation between field strength and
electric field contains such a factor,
$F_{\mu\nu}=2E\,\epsilon_{\mu\nu}$. Thus, apart from the sign, the
normalization in \eqref{eq:Gauss} leads to the standard normalization
of electric charge in 2D. The sign is a consequence of our desire to
have positive $E$ for positive $q_e$ in the case of negative
$\alpha$.} The Gauss law \eqref{eq:Gauss} allows to rewrite the gauge
shift of the action \eqref{eq:bm30} in a suggestive way:
\eq{ \de_\Lambda I = \de_\Lambda I_{\rm new} = 2\pi\,q_e \,q_m~.
}{eq:angelinajolie}
Thus, as long as magnetic and electric charge obey the Dirac
quantization condition \eq{ q_e \,q_m \in\mathbb{Z}~, }{eq:bm32} the
action just shifts by multiples of $2\pi$. We shall assume that this
is the case.  Then $I_{\rm new}$ and $I$ are gauge invariant modulo
$2\pi$ despite of the apparent gauge non-invariance of the boundary
mass term $m \,A^a A_a$.

In conclusion, the full action \eqref{Action} is gauge invariant with
respect to all gauge variations \eqref{eq:bm27} that preserve the
boundary conditions \eqref{eq:FG} provided the integral in
\eqref{eq:bm28} vanishes (modulo $2\pi$). This is the case if the
Dirac quantization condition \eqref{eq:bm32} holds.

\section{Boundary Stress Tensor and Central Charge}
\label{sec:BSTcc}

The behavior of the on-shell action is characterized by the linear
response functions of the boundary theory\,\footnote{These are the
same conventions as in \cite{Balasubramanian:1999re}. The boundary
current and stress tensor used here is related to the definitions in
\cite{Hartman:2008dq} by $J^{a}={1\over 2\pi} J^{a}_\textrm{\tiny
HS}$ and $T^{ab}={1\over 2\pi}T^{ab}_\textrm{\tiny HS}$. }
\begin{align}\label{2dTJ}
T^{ab} = \frac{2}{\sqrt{-h}}\,\frac{\delta I}{\delta h_{ab}}~, \quad
\quad \quad J^{a} = \frac{1}{\sqrt{-h}}\,\frac{\delta I}{\delta
A_{a}} ~.
\end{align}
The response function for the dilaton, which is not relevant for the
present considerations, is discussed in \cite{Grumiller:2007ju}. The
general expressions \eqref{hmomentum} and \eqref{Amomentum} give
\begin{subequations}\label{tensorcurrent}
\begin{align}\label{stresstensor}
T_{tt} & \, =  \frac{\alpha}{\pi}\,\big( - \frac{2}{L} \,
h_{tt}\,e^{-2 \phi}  - \frac{L}{2} \,A_{t} A_{t}\big)~, \\
\label{bndycurrent} J^{t} & \, = \frac{\alpha}{2\pi}\,\big(
-L^{2}n_\mu F^{\mu t} + 2LA^{t} \big) ~.
\end{align}
\end{subequations}
We want to find the transformation properties of these functions
under the asymptotic symmetries of the theory; {\it i.e.} under the
combination of bulk diffeomorphisms and $U(1)$ gauge transformations
that act non-trivially at $\dM$, while preserving the boundary
conditions and the choice of gauge.

A diffeomorphism $x^{\mu} \to x^{\mu} + \epsilon^{\mu}(x)$ transforms the fields as
\begin{align}
\delta_{\epsilon} g_{\mu\nu} = \nabla_{\mu} \epsilon_{\nu} + \nabla_{\nu} \epsilon_{\mu}~, \quad \quad \quad \delta_{\epsilon} A_{\mu} = \epsilon^{\nu} \, \nabla_{\nu} A_{\mu} + A_{\nu} \nabla_{\mu} \epsilon^{\nu} ~.
\end{align}
The background geometry is specified by the gauge conditions
$g_{\eta\eta}=1$, $g_{\eta t}=0$, and the boundary condition that
fixes the leading term $g_{tt}^{(0)}$ in the asymptotic expansion
\eqref{eq:FGmetric} of $h_{tt}$. These conditions are preserved by
the diffeomorphisms
\begin{equation} \label{diffeo}
\epsilon^{\eta} = - \frac{L}{2}\,\partial_{t} \xi(t)~, \quad \quad \epsilon^{t} = \xi(t) +
{L^2\over2}\left(  e^{4 \eta/L} + h_1(t) \right)^{-1}\,\partial_{t}^{\,2}\xi(t)~,
\end{equation}
where $\xi$ is an arbitrary function of the coordinate $t$. Under
\eqref{diffeo}, the boundary metric transforms according to
\begin{align}\label{metrictransform}
\delta_{\epsilon} h_{tt} & \, = -\left(1 + e^{-4\eta/L}h_1(t) \right)\,\left( h_1(t) \,\partial_{t} \xi(t) + \frac{1}{2}\, \xi(t) \partial_t h_1(t) + \frac{L^{\,2}}{4}\, \partial_{t}^{3} \xi(t)\right) ~.
\end{align}
Turning to the gauge field, the change in $A_{\eta}$ due to the diffeomorphism
\eqref{diffeo} is
\begin{align}
\delta_{\epsilon} A_{\eta} & \, = -2\,e^{-\phi}\,\frac{\left(e^{2\eta/L}-h_1(t)\,e^{-2\eta/L} \right)}{\left(e^{2\eta/L} + h_1(t)\,e^{-2\eta/L} \right)^2}\,\partial_{t}^{2}\xi(t) ~.
\end{align}
Thus, diffeomorphisms with $\partial_{t}^{2}\xi \neq 0$ do not
preserve the $U(1)$ gauge condition $A_{\eta} = 0$. The gauge is
restored by the compensating gauge transformation $A_{\mu} \to A_{\mu} +
\partial_{\mu} \Lambda$, with $\Lambda$ given by
\begin{gather}\label{compensating}
\Lambda = - L
\,e^{-\phi}\,\left(e^{2\eta/L}+h_1(t)\,e^{-2\eta/L}\right)^{-1}\partial_{t}^{2}\xi(t)
~.
\end{gather}
The effect of the combined diffeomorphism and $U(1)$ gauge transformation on $A_{t}$
is
\begin{equation}\label{Atransform}
\left(\delta_{\epsilon} + \delta_{\Lambda}\right) A_{t} = -
e^{-2\eta/L}\,e^{-\phi}\,\left( {1\over
L}\xi(t) \partial_{t} h_1(t) + \frac{2}{L}\,h_1(t)\,\partial_{t} \xi(t) +
\frac{L}{2}\,\partial_{t}^{3} \xi(t) \right) ~.
\end{equation}
This transformation preserves the boundary condition \eqref{eq:FGgf}
for $A_{t}$, as well as the condition $A_t^{(1)}=0$ that was used to
fix the residual $U(1)$ gauge freedom. Thus, the asymptotic
symmetries of the theory are generated by a diffeomorphism
\eqref{diffeo} accompanied by the $U(1)$ gauge transformation
\eqref{compensating}. Under such transformations the metric and
gauge field behave as \eqref{metrictransform} and
\eqref{Atransform}, respectively.

We can now return to our goal of computing the transformation of the linear response
functions \eqref{tensorcurrent} under the asymptotic symmetries of the theory.
The change in the stress tensor \eqref{stresstensor} due to the
combined diffeomorphism \eqref{diffeo} and $U(1)$ gauge transformation \eqref{compensating}
takes the form
\begin{equation}\label{deltaTtt}
(\delta_{\epsilon}+\delta_\Lambda) T_{tt} = 2\,T_{tt}\,\partial_{t} \xi + \xi \,
\partial_{t} T_{tt} - \frac{c}{24 \pi}\,L\,\partial_{t}^{3}\xi(t)
~.
\end{equation}
The first two terms are the usual tensor transformation due to the diffeomorphism. In addition, there is an anomalous term generated by the $U(1)$ component of the asymptotic symmetry.
We included a factor $L$ in the anomalous term in \eqref{deltaTtt} in order to make the central charge $c$ dimensionless.
Using the expressions \eqref{metrictransform} and \eqref{Atransform} for the transformation of the fields
we verify the general form \eqref{deltaTtt} and determine the central charge
\begin{equation}\label{centcharge1}
c = -24 \,\alpha e^{-2\phi} ~.
\end{equation}
The relation \eqref{eq:G2} allows us to rewrite \eqref{centcharge1}
in the more aesthetically pleasing form
\begin{equation}\label{centcharge}
c = \frac{3}{G_2}~.
\end{equation}
The requirement that the central charge should be positive determines
$\alpha<0$ as the physically correct sign. We shall see the same
(unusual) sign appearing as the physically correct one in later
sections.

Another suggestive expression for the central charge is
\begin{equation}
c = 3\, {\rm Vol}_L \,{\cal L}_{\rm 2D}~, \label{funnyc}
\end{equation}
where the volume element ${\rm Vol}_L=2\pi L^2$ and Lagrangian
density ${\cal L}_{\rm 2D}= {4\alpha\over\pi L^2}e^{-2\phi}$ is related
to the on-shell bulk action \eqref{eq:aa} by
\begin{equation}
I_{\rm bulk}\big|_{\textrm{\tiny EOM}}=-\int_{\cal M}
d^2x\sqrt{-g}\,{\cal L}_{\rm 2D}~.
\end{equation}
The central charge \eqref{funnyc} is the natural starting point for
computation of higher derivative corrections to the central charge,
in the spirit of
\cite{Kraus:2005vz,Sen:2005wa,Gupta:2008ki,Sen:2008vm}.

So far we considered just the transformation property of the energy momentum
tensor \eqref{stresstensor}. We should also consider the response of the
boundary current \eqref{bndycurrent} to a gauge
transformation. Generally, we write the transformation of a current as
\begin{gather}
\delta_{\Lambda} J_{t} = -{k\over4\pi}\,L \,\partial_{t}\Lambda~,
\label{eq:Jtrafo}
\end{gather}
where the level $k$ parametrizes the gauge anomaly. The only term in \eqref{bndycurrent}
that changes under a gauge transformation is the term proportional to $A_{t}$.
The resulting variation of the boundary current takes the form \eqref{eq:Jtrafo}
with the level
\begin{equation}
k = -{4\alpha} = {1\over 2G_2} e^{2\phi}~.
\label{eq:level}
\end{equation}


Our definitions of central charge \eqref{deltaTtt} and level in
\eqref{eq:Jtrafo} are similar to the corresponding definitions in 2D
CFT. However, they differ by the introduction of the AdS scale $L$,
needed to keep these quantities dimensionless. We could have
introduced another length scale instead, and the anomalies would then
be rescaled correspondingly as a result. Since $c$ and $k$ would
change the same way under such a rescaling we may want to express the
central charge \eqref{centcharge} in terms of the level
\eqref{eq:level} as
\begin{gather}\label{centralcharge1}
c =  6\,k\,e^{-2 \phi}\,.
\end{gather}
This result is insensitive to the length scale introduced in the definitions of the anomalies,
as long as the same scale is used in the two definitions.

Expressing the dilaton \eqref{eq:dilaton} in terms of the electric
field we find yet another form of the central charge
\begin{equation}
c = \frac{3}{2}\,k E^{2}L^{4} ~.
\label{centralcharge}
\end{equation}
As it stands, this result is twice as large as the result found in
\cite{Hartman:2008dq}. However, there the anomaly is attributed to
two contributions, from $T_{++}$ and $T_{--}$ related to the two
boundaries of global AdS$_2$. We introduce a single energy-momentum
tensor $T_{tt}$, as seems appropriate when the boundary theory has
just one spacetime dimension. In general spacetimes, $T_{tt}$ would
be a density but in one spacetime dimension there are no spatial
dimensions, and so the ``density" is the same as the energy. Such an
energy-momentum tensor cannot be divided into left- and right-moving
parts. Thus our computation agree with \cite{Hartman:2008dq} even
though our interpretations differ.

\section{3D Reduction and Connection with 2D}
\label{sec:5}
Asymptotically AdS$_2$ backgrounds have a non-trivial $SL(2,\mathbb{R})$
group acting on the boundary that can be interpreted as one of the
two $SL(2,\mathbb{R})$ groups associated to AdS$_3$. To do so, we compactify
pure gravity in 3D with a negative cosmological constant on a circle
and find the map to the Maxwell-dilaton gravity \eqref{eq:aa}. This
dimensional reduction also shows that the AdS$_2$ boundary
stress tensor and central charge found in this paper are consistent
with the corresponding quantities in AdS$_3$.

\subsection{Three dimensional gravity}

Our starting point is pure three dimensional gravity described by an action
\begin{equation}\label{3dbulk}
I={1\over 16\pi G_3}\int d^3x
\sqrt{-g}\,\left(\Rthree+{2\over\ell^2}\right)+{1\over8\pi G_3}\int
d^2y\sqrt{-{\gamma}}\,\left(\Kthree-{1\over\ell}\right)~,
\end{equation}
that is a sum over bulk and boundary actions like in the schematic equation \eqref{eq:bulkandboundary}.
The 3D stress-tensor defined as
\begin{equation}\label{emtensor}
\delta I={1\over 2}\int d^2y\sqrt{-\gamma}~T^{ab}_{\textrm{\tiny
3D}}~\delta\gamma_{ab}~,
\end{equation}
becomes \cite{Balasubramanian:1999re}
\begin{equation}\label{emtens}
T_{ab}^{\textrm{\tiny 3D}}=-{1\over8\pi
G_3}\Big(\Kthree_{ab}-\Kthree\gamma_{ab}+{1\over
\ell}\gamma_{ab}\Big)~.
\end{equation}

For asymptotically AdS$_3$ spaces we can always choose
Fefferman-Graham coordinates, where the bulk metric takes the form
\begin{equation}\label{FG}
ds^2=d\eta^2+\gamma_{ab}\,dy^ady^b~, \quad
\gamma_{ab}=e^{2\eta/\ell}\gamma^{(0)}_{ab}+\gamma^{(2)}_{ab}+\ldots~.
\end{equation}
The functions $\gamma^{(i)}_{ab}$ depend only on the boundary
coordinate $y^a$ with $a,b=1,2$. The boundary is located at
$\eta\to\infty$, and $\gamma^{(0)}_{ab}$ is the 2D boundary metric
defined up to conformal transformations. The energy momentum tensor
\eqref{emtens} evaluated in the coordinates \eqref{FG} is
\begin{equation}\label{emtenso}
T^{\textrm{\tiny 3D}}_{ab}={1\over 8\pi G_3\ell}\Big( \gamma^{(2)}_{ab} -
\gamma_{(0)}^{cd} \gamma^{(2)}_{cd} \gamma^{(0)}_{ab} \Big)~.
\end{equation}

\newcommand{\FGA}{g_+}
\newcommand{\bFGA}{g_-}

In the case of pure gravity \eqref{3dbulk} we can be more explicit
and write the exact solution as \cite{Skenderis:1999nb}
\begin{equation}\label{FGAdS3}
ds^2=d\eta^2+\Big({\ell^2\over4}e^{2\eta/\ell}+4\FGA\bFGA e^{-2\eta/\ell}\Big)dx^+dx^- + \ell\Big(\FGA(dx^+)^2+\bFGA(dx^-)^2\Big)~.
\end{equation}
We assumed a flat boundary metric $\gamma^{(0)}_{ab}$ parameterized
by light-cone coordinates $x^\pm$. The function $\FGA$ ($\bFGA$)
depends exclusively on $x^+$ ($x^-$). For this family of solutions
the energy-momentum tensor \eqref{emtenso} becomes
\begin{equation}\label{emtenthree}
T^{\textrm{\tiny 3D}}_{++}={1\over 8\pi G_3}\,\FGA~, \qquad
T^{\textrm{\tiny 3D}}_{--}={1\over 8\pi G_3}\,\bFGA~.
\end{equation}

\subsection{Kaluza-Klein reduction}\label{sec:4.2}

Dimensional reduction is implemented by writing the 3D metric as
\begin{equation}\label{KKmetric}
ds^2=e^{-2\psi}\ell^2(dz+\tilde{A}_\mu dx^\mu)^2+\tilde{g}_{\mu\nu}dx^\mu
dx^{\nu}~.
\end{equation}
The 2D metric $\tilde{g}_{\mu\nu}$, the scalar field ${\psi}$, and
the gauge field $\tilde{A}_\mu$ all depend only on
$x^\mu$ ($\mu=1,2$). The coordinate $z$ has period $2\pi$. The 3D
Ricci scalar expressed in terms of 2D fields reads
\begin{equation}\label{3dRicci}
\Rthree=\tilde{R}-2e^{\psi}\tilde{\nabla}^2e^{-\psi}-{\ell^2\over4}
e^{-2\psi}\tilde{F}^2~.
\end{equation}
The 2D scalar curvature $\tilde{R}$,
and the covariant derivatives $\tilde{\nabla}_\mu$, are constructed from
$\tilde{g}_{\mu\nu}$. Inserting (\ref{3dRicci}) in the 3D bulk
action in (\ref{3dbulk}) gives the 2D bulk action
\begin{equation}\label{2dKKbulk}
\tilde{I}_{\rm bulk}={\ell\over 8 G_3}\int d^2x
\sqrt{-\tilde{g}}e^{-\psi}\Big(\tilde{R}+{2\over\ell^2}-{\ell^2\over4}e^{-2\psi}\tilde{F}^2\Big)~.
\end{equation}

The action (\ref{2dKKbulk}) is on-shell equivalent  to the action
\eqref{Action} for the constant dilaton solutions \eqref{Solution}. To find the precise dictionary
we first compare the
equations of motion. Variation of the action (\ref{2dKKbulk}) with
respect to the scalar $\psi$ and metric $\tilde{g}_{\mu\nu}$ gives
\begin{subequations}
\begin{align}
\tilde{R}+{2\over\ell^2}-{3\ell^2\over4}\,e^{-2\psi}\tilde{F}^2&=0~,\label{eompsi}\\
\tilde{g}_{\mu\nu}\Big({1\over\ell^2}-{\ell^2\over8}\,e^{-2\psi}\tilde{F}^2\Big)
+{\ell^2\over2}\,e^{-2\psi}\tilde{F}_{\mu\alpha}\tilde{F}_{\nu}^{~\alpha}&=0~,\label{eomKK}
\end{align}
\end{subequations}
which implies\footnote{A check on the algebra: inserting  \eqref{eomKK1}
into the formula \eqref{3dRicci} for the 3D Ricci scalar yields $\Rthree=-6/\ell^2$, concurrent with our definition of the 3D AdS radius.}
\begin{subequations}\label{eomKK1}
\begin{align}
e^{-2\psi}\tilde{F}^2 & =-{8\over\ell^4}~,\label{FKK} \\
\tilde{R}&=-{8\over\ell^2}~.\label{RicciKK}
\end{align}
\end{subequations}
The analogous equations derived from the 2D action \eqref{redEqns}
take the same form, but with the identifications
\begin{subequations}\label{dicc}
\begin{align}
\tilde{g}_{\mu\nu}&=a^2 g_{\mu\nu}~,\label{diccmectric}\\
\ell&=aL~,\label{eq:laL}\\
e^{-\psi}\tilde{F}_{\mu\nu}&={1\over2}e^{\phi}F_{\mu\nu}~,\label{diccgauge}
\end{align}
\end{subequations}
with $a$ an arbitrary constant.

In order to match the overall normalization on-shell  we evaluate
the bulk action (\ref{2dKKbulk}) using the on-shell relations
(\ref{FKK}) and (\ref{RicciKK})
\begin{equation}
\tilde{I}_{\rm bulk}=-{\ell\over2G_3}\int d^2x{\sqrt{-\tilde{g}}\over\ell^2}{
e^{-\psi}}~.
\end{equation}
and compare with the analogous expression
\begin{equation}
{I}_{\rm bulk}={4\alpha\over\pi}\int d^2x{\sqrt{-{g}}\over L^2}{e^{-2\phi}}~.
\label{onshellact}
\end{equation}
computed directly from the 2D action \eqref{Action}. Equating the on-shell actions
$I_{\rm bulk}=\tilde{I}_{\rm bulk}$ and simplifying using (\ref{diccmectric}), \eqref{eq:laL} we find
\begin{equation}\label{diccscalar}
\alpha = -\frac{\pi\,\ell}{8G_3} e^{2\phi-\psi}\,.
\end{equation}
We see again that the unusual sign $\alpha<0$ is the physically correct one.
According to \eqref{eq:G2} we can write the 3D/2D identification as
\begin{equation}
{1\over G_2} = \frac{\pi\ell e^{-\psi}}{G_3}~.
\end{equation}

So far we determined the 3D/2D on-shell dictionary by comparing
equations of motions and the bulk action. In appendix \ref{app:A} we
verify that the same identification \eqref{diccscalar} also
guarantees that the boundary actions agree. Additionally, we show
that the 3D/2D dictionary identifies the 3D solutions \eqref{FGAdS3}
with the general 2D solutions \eqref{Solution}. These checks give
confidence in our 3D/2D map.

In summary, our final result for the dictionary between the 2D theory
and the KK reduction of the 3D theory is given by the identifications
\eqref{dicc} and the relation \eqref{diccscalar} between
normalization constants. We emphasize that the map is on-shell; it is
between solutions and their properties. The full off-shell theories
do not agree, as is evident from the sign in \eqref{diccscalar}. The
restriction to on-shell configurations will not play any role in this
paper but it may be important in other applications.

\subsection{Conserved currents and central charge}

Applying the 3D/2D dictionary from the previous subsection (and
elaborations in appendix \ref{app:A}), we now compare the linear
response functions and the central charge computed by reduction from
3D to those computed directly in 2D.

The starting point is the 3D energy momentum tensor (\ref{emtensor}).
The KK-reduction formula (\ref{KKmetric2}) decomposes the variation of
the boundary metric $\gamma_{ab}$ as
\begin{equation}\label{delta}
\delta\gamma_{ab}= \left(
\begin{array}{cc}
1 ~& 0 \\
0 ~& 0 \\
\end{array}
\right)\delta \tilde{h}_{tt} + \ell^2e^{-2\psi}\left(
\begin{array}{cc}
2\tilde{A}_t & 1 \\
1 & 0 \\
\end{array}
\right)\delta \tilde{A}_t~,
\end{equation}
and the determinant $\sqrt{-\gamma}=\ell
e^{-\psi}\sqrt{-\tilde{h}}$ so that the 3D stress tensor (\ref{emtensor}) becomes
\begin{eqnarray}\label{deltaI}
\delta I &= &\int dx \sqrt{-\tilde{h}} \Big[ {1\over2}\big(2\pi\ell
e^{-\psi}\big)T^{tt}_{3d}~\delta \tilde{h}_{tt}+ \big(\pi\ell^3
e^{-3\psi}\big)\big(T^{tt}_{3d}\tilde{A}_{t}+T^{zt}_{3d}\big)]~2\delta
\tilde{A}_{t}\Big]~\cr \label{delta2d}
 & =&\int dx \sqrt{-{h}}(h^{tt})^2 \Big[ {1\over2}\big(2\pi L
e^{-\psi}T_{tt}^{3d}\big)\delta {h}_{tt} + \big({\pi\over2} L^3
e^{-\psi+2\phi}{A}_{t}T_{tt}^{3d}\big)\delta A_{t}\Big] ~,
\end{eqnarray}
where we used the 3D-2D dictionary (\ref{dicc}) and wrote the
variation of the boundary fields as
\begin{equation}\label{delta1}
\delta \tilde{h}_{tt}=a^2\delta h_{tt}~,\quad \delta
\tilde{A}_{t}={1\over2}e^{\psi+\phi}\delta A_t~.
\end{equation}
Indices of the stress tensor in (\ref{deltaI}) are lowered and raised
with $\tilde{h}_{tt}$ and $\tilde{h}^{tt}$, respectively. Comparing
(\ref{delta2d}) with the 2D definition of stress tensor and current
\eqref{2dTJ} we find
\begin{subequations}\label{Stress3d2d}
\begin{align}
T^{\textrm{\tiny 2D}}_{tt}&=2\pi L e^{-\psi}T_{tt}^{\textrm{\tiny 3D}}~,\label{stressKK2d}\\
J_{t}&={\pi\over2} L^3
h^{tt}e^{-\psi+2\phi}{A}_{t}T_{tt}^{\textrm{\tiny 3D}}~,
\end{align}
\end{subequations}
for the relation between 3D and 2D quantities.

The next step is to rewrite the 3D energy momentum tensor
\eqref{emtenthree} in a notation more appropriate for comparison with
2D. We first rescale coordinates according to \eqref{tcoordef} and
then transform into 2D variables using \eqref{psidiv},
\eqref{ctwodic}. The result is
\begin{equation}\label{stressKK}
T^{\textrm{\tiny 3D}}_{tt}=-{1\over 8\pi G_3\ell}h_1~, \qquad
T^{\textrm{\tiny 3D}}_{zz}={1\over 8\pi G_3}\ell e^{-2\psi}~.
\end{equation}
Inserting these expressions in (\ref{Stress3d2d}), along with the
asymptotic values of the background fields in the solution
\eqref{Solution}, we find
\begin{subequations}\label{emKK2d}
\begin{align}
T^{\textrm{\tiny 2D}}_{tt}&={2\alpha\over L\pi}e^{-2\phi}\,h_1~,\label{emKKstress}\\
J_{t}&=-{2\alpha\over \pi}e^{-\phi}e^{-2\eta/L}\,h_1~,\label{emKKcurrent}
\end{align}
\end{subequations}
after simplifications using our 3D-2D dictionary \eqref{dicc} and
the rescaling mentioned just before \eqref{h1}. The current \eqref{emKKcurrent}
vanishes on the boundary $\eta\to\infty$ but the subleading term given here
is significant for some applications. The expressions \eqref{emKK2d} are
our results for the 2D linear response functions, computed by
reduction from 3D. They should be compared with the analogous
functions \eqref{tensorcurrent} defined directly in 2D, with those
latter expressions evaluated on the solution \eqref{Solution}. These
results agree precisely.

Using the relations between the conserved currents, we now proceed
to compare the central charges in 2D and 3D. Under the
diffeomorphisms which preserve the three dimensional boundary, the
3D stress tensor transforms as \cite{Balasubramanian:1999re}
\begin{equation}\label{delta3d}
\delta T_{tt}^{\textrm{\tiny 3D}}=2T_{tt}^{\textrm{\tiny
3D}}\partial_t\xi(t)+\xi(t)\partial_tT_{tt}^{\textrm{\tiny
3D}}-{c\over24\pi}\partial^3_t\xi(t)~,
\end{equation}
with the central term given by the standard Brown--Henneaux central
charge
\begin{equation}\label{BrownHenn}
c_{\textrm{\tiny 3D}}={3\ell\over 2G_3}~.
\end{equation}
 From the relation \eqref{stressKK2d} between 2D and 3D stress
tensor, and by comparing the transformations \eqref{deltaTtt} and
\eqref{delta3d}, the central charges are related as
\begin{equation}\label{c2c3}
c_{\textrm{\tiny 2D}}=2\pi e^{-\psi} c_{\textrm{\tiny 3D}}~.
\end{equation}
Inserting \eqref{BrownHenn} and using \eqref{diccscalar} we find
\begin{equation}\label{c2}
c_{\textrm{\tiny 2D}}=2\pi e^{-\psi} \left({ 3\ell\over
2G_3}\right) = -24\alpha e^{-2\phi}~.
\end{equation}
This is the result for the 2D central charge, obtained by reduction
from 3D. It agrees precisely with the central charge
\eqref{centcharge1} obtained directly in 2D.

In summary, in this section we have given an explicit map between 3D
and 2D. We have shown that it correctly maps the equations of motion
and the on-shell actions, it maps 3D solutions to those found
directly in 2D, it maps the linear response functions correctly
between the two pictures, and it maps the central charge correctly.

\section{Black Hole Thermodynamics}
\label{sec:6}

In this section we apply our results to discuss the entropy of 2D
black holes. We start by computing the temperature and mass of the
black hole and the relation of these quantities to the 2D stress
tensor. By using the renormalized on-shell action and the first law
of thermodynamics, we obtain the Bekenstein-Hawking entropy. Finally,
we discuss the identification of the black hole entropy with the
ground state entropy of the dual CFT.

\subsection{Stress tensor for AdS$_2$ black holes}
For $h_0=1$ and constant $h_1$, the solution \eqref{Solution} becomes
\begin{subequations}\label{BH2d}
\begin{align}
ds^2 &= d\eta^2-{1\over4}e^{4\eta/L}\left(1+h_1e^{-4\eta/L}\right)^2dt^2~,
\\
A_t &= {1\over
L}e^{-\phi}e^{2\eta/L}\left(1-h_1e^{-4\eta/L}\right)~.
\end{align}
\end{subequations}
Solutions with positive $h_1$ correspond to global AdS$_2$ with
radius $\ell_A=L/2$, while solutions with negative $h_1$ describe
black hole geometries.

An AdS$_2$ black hole with horizon at $\eta=\eta_0$ corresponds to
$h_1$ given by
\begin{equation}\label{Horizon}
h_1 = - e^{4\eta_0/L}.
\end{equation}
Regularity of the Euclidean metrics near the horizon determines the imaginary periodicity
$t\sim t+i\beta$ as
\begin{equation}\label{beta}
\beta=\pi Le^{-2\eta_0/L}~.
\end{equation}
We identify the temperature of the 2D black hole as $T=\beta^{-1}$.

Our general AdS$_2$ stress tensor \eqref{tensorcurrent}, \eqref{emKKstress} is
\begin{equation}
T_{tt}={2\alpha\over \pi L}e^{-2\phi} h_1 = - {h_1\over 4\pi G_2 L}~.
\label{adstensor}
\end{equation}
The stress tensor for global AdS$_2$ ($h_1>0$) is negative. This is reasonable, because
the Casimir energy of AdS$_3$ is negative as well. Importantly, the black hole solutions
($h_1<0$) are assigned positive energy, as they should be. The assignment $\alpha<0$
is needed in \eqref{adstensor}
to reach this result, giving further confidence in our determination of that sign.

We can rewrite the stress tensor \eqref{adstensor} as
\begin{equation}
T_{tt} = {\pi LT^2\over 4G_2}= c\, {\pi LT^2\over 12}~,
\end{equation}
where we used the central charge \eqref{centcharge}. We interpret this form of the
energy as a remnant of the 3D origin of the theory, as the right movers of a 2D CFT.

The mass is generally identified as the local charge of the current
generated by the Killing vector $\partial_t$. This amounts to the
prescription
\begin{equation}
M  = \sqrt{-g^{tt}}\, T_{tt} = 2e^{-2\eta/L}\,T_{tt} \to 0~,
\label{eq:admmass}
\end{equation}
for the mass measured asymptotically as $\eta\,\to\,\infty$. The
solutions \eqref{BH2d} are therefore all assigned vanishing mass, due
to the redshift as the boundary is approached. We will see in the
following that this result is needed to uphold the Bekenstein-Hawking
area law.

\subsection{On-shell action and Bekenstein-Hawking entropy}
The boundary terms in \eqref{Action} were constructed so that the variational principle
is well-defined, but they are also supposed to cancel divergences and render the
on-shell action finite. It is instructive to compute its value.

The on-shell bulk action \eqref{onshellact} becomes
\begin{align}\label{Ibulk}
I_{\rm bulk}&={2\alpha\over\pi L^2}e^{-2\phi} \int dt d\eta\,
\left(e^{2\eta/L}+h_1e^{-2\eta/L}\right) \nonumber\\
&={\alpha\beta\over \pi L}e^{-2\phi}
\left.\left(e^{2\eta/L}-h_1e^{-2\eta/L}\right)\right|_{\eta_0}^{\infty}~,
\end{align}
for the 2D black hole \eqref{BH2d}. The boundary terms in
\eqref{Action} were evaluated in \eqref{onshellct} with the result
\begin{equation}
I_{\rm boundary} = - {2\alpha\beta\over \pi L} e^{-2\phi} \sqrt{-h_{tt}} =
- {\alpha\beta\over \pi L} e^{-2\phi} e^{2\eta/L}~.
\end{equation}
The divergence at the boundary $\eta\to\infty$ cancels the
corresponding divergence in the bulk action \eqref{Ibulk}. The
renormalized on-shell action becomes finite with the value
\begin{equation}\label{Ionshell}
I = I_{\rm bulk} + I_{\rm boundary} =- {2\alpha\beta\over \pi L} e^{-2\phi} e^{2\eta_0/L}  = -2\alpha e^{-2\phi}={1\over 4G_2}~.
\end{equation}
The third equality used \eqref{beta} and the last one used \eqref{eq:G2}.

We computed the on-shell action in Lorentzian signature to conform
with the conventions elsewhere in the paper. The Euclidean action has
the opposite sign $I_E = -I$, and it is that action which is related
to the free energy in the standard manner
\begin{equation}
\beta F  = I_E = \beta M - S~,
\end{equation}
when we consider the canonical ensemble.\footnote{Strictly, the
on-shell action is related to a thermodynamic potential that is a
function of the temperature $T$ and the electrostatic potential
$\Phi$. However, the boundary term for the gauge field leads to a net
charge $Q = 0$, and so the thermodynamic potential reduces to the
standard Helmholtz free energy.} We found vanishing $M$ in
\eqref{eq:admmass} and so the black hole entropy becomes
\begin{equation}\label{Sthermo}
S= - I_E =  I = {1\over 4G_2}~.
\end{equation}
This is the standard Bekenstein-Hawking result.

\subsection{Black hole entropy from Cardy's formula}

One of the motivations for determining the central charge of AdS$_2$
is that it may provide a short-cut to the black hole entropy. We will
just make preliminary comments on this application.

A 2D chiral CFT with $c_0$ degrees of freedom living on a circle with
radius $R$ has entropy given by the Cardy formula
\begin{equation}
S = 2\pi \sqrt{{c_0\over 6}(2\pi R H)}~.
\label{eq:cardy}
\end{equation}
Here $H$ denotes the energy of the system. This formula generally
applies when $2\pi RH\gg {c_0\over 24}$ but, for the CFTs relevant
for black holes, we expect it to hold also for $2\pi RH\sim {c_0\over
24}$ \cite{Dijkgraaf:2000fq}. Since the Casimir energy for such a
theory is $2\pi RH= {c_0\over 24}$ we recover the universal ground
state entropy
\begin{equation}
S = 2\pi \cdot {c_0\over 12}~.
\end{equation}
Relating the number of degrees of freedom $c_0$ to our result for the
central charge \eqref{centcharge} as $c_0 = c/(2\pi)$ we find the
ground state entropy
\begin{equation}
S = {c\over 12} = {1\over 4G_2}~,
\label{bekhawk}
\end{equation}
in agreement with the Bekenstein-Hawking entropy.

The relation $c_0 = c/(2\pi)$ is not self-evident. We have defined
the central charge by the transformation property \eqref{deltaTtt}
with stress tensor normalized as in \eqref{2dTJ}. This gives the same
normalization of central charge as in \cite{Hartman:2008dq}. As we
have already emphasized, the length scale $L$ introduced in
\eqref{deltaTtt} to render the central charge dimensionless is rather
arbitrary. We could have introduced $2\pi L$ instead, corresponding
to
\begin{equation}\label{czerodef}
(\delta_{\epsilon}+\delta_\Lambda) T_{tt} = 2\,T_{tt}\,\partial_{t} \xi + \xi \,
\partial_{t} T_{tt} - \frac{c_0}{12}\,L\,\partial_{t}^{3}\xi(t)~.
\end{equation}
It is apparently this definition that leads to $c_0$, the measure of degrees of
freedom.

The situation is illuminated by our 3D-2D dictionary. We can
implement this by using the 3D origin of the 2D coordinate $t$
\eqref{tcoordef} or, simpler, the 3D origin of the 2D central charge
\eqref{c2c3}. In 3D the dimensionless central charge that counts the
degrees of freedom is introduced without need of an arbitrary scale.
The relation \eqref{c2c3} to the 2D central charge therefore
motivates the factor $2\pi$ in $c_0 = c/(2\pi)$. Furthermore, there
is a conformal rescaling of the central charge due to an induced
dilaton $e^{-\psi}$. To get a feel for this consider the canonical 4D
BPS black holes \cite{Klebanov:1996mh,Balasubramanian:1996rx},
supported by four mutually BPS charges $n_1, n_2, n_3, n_0$ of which
$n_0$ is the KK-momentum along the circle. The conformal rescaling
brings the 3D central charge $c_{3D} = 6n_1 n_2 n_3$ into the more
symmetrical value
\begin{equation}
c_0=6\sqrt{n_1 n_2 n_3 n_0}
\end{equation}
It would be interesting to understand this value directly from the 2D
point of view.

It is natural to consider a more general problem. The 2D black holes
\eqref{BH2d} are lifted by our 2D/3D map in section \S\ref{sec:5} to
the general BTZ black holes in three dimensions. The BTZ black holes
are dual to a 2D CFT, with both right and left movers. The 2D
description keeps only one chirality and so it is challenging to
understand how the general entropy can be accounted for directly in
2D. Our result equating the ground state entropy of the chiral 2D CFT
with the Bekenstein-Hawking entropy of any AdS$_2$ black hole
indicates that this is in fact possible, but the details remain
puzzling.

\section{Backgrounds with Non-constant Dilaton}\label{sec:new}

In this section we generalize our considerations to backgrounds with
non-constant dilaton. We find that the counterterms determined for
constant dilaton give a well-defined variational principle also in
the case of a non-constant dilaton. We discuss some properties of the
general solutions. In particular we identify an extremal solution
that reduces to the constant dilaton solution \eqref{Solution} in a
near horizon limit. For recent work on non-constant dilaton solutions
in 2D Maxwell-Dilaton gravity see \cite{Cadoni:2008mw}.

\subsection{General solution with non-constant dilaton}

We start by finding the solutions to the equations of motion. The
spacetime and gauge curvature are determined by solving
$\EE_{\phi}=0$ and $\EE_\mu=0$ in \eqref{eq:eom}, which gives
\begin{equation}
R=-\frac{8}{L^2}\,,\qquad F_{\mu\nu} = 2E \,\epsilon_{\mu\nu}~.
\label{eq:gen1}
\end{equation}
In the case of non-constant dilaton
we may use the dilaton as one of the coordinates
\begin{subequations}
\label{eq:lindil}
\begin{equation}
e^{-2\phi} = \frac{r}{L}~. \label{eq:gen2}
\end{equation}
This statement is true everywhere except on bifurcation points of
bifurcate Killing horizons. We do not exhaustively discuss global
issues here and therefore disregard this subtlety. For dimensional
reasons we have included a factor $1/L$ on the right hand side of
the definition \eqref{eq:gen2}. Using the residual gauge freedom we
employ again a gauge where the line element is diagonal and the
gauge field has only a time component,
\begin{equation}
ds^2 = g_{rr}\,dr^2 + g_{tt}\, dt^2\,,\qquad A_\mu dx^\mu = A_t \,dt\,.
\label{eq:gen3}
\end{equation}
Solving $\EE_{\phi}=0$ yields $g_{tt}=-1/g_{rr}$, and the last
equations of motion $\EE_{\mu\nu}=0$ gives
\begin{equation}
g_{tt} = - \frac{4r^2}{L^2} + 2 L^3 E^2\, r + 4M~, \label{eq:gen4}
\end{equation}
and
\begin{equation}
A_t = 2E r\,.
\label{eq:gen5}
\end{equation}
\end{subequations}
The electric field $E$ and `mass' $M$ are constants of motion. The
former has dimension of inverse length squared, the latter is
dimensionless in our notation.

There is a Killing vector $k={\partial_t}$ that leaves the metric,
gauge field and dilaton invariant. There are two other Killing
vectors that leave invariant the metric, but not the dilaton. This is
the breaking of $SL(2,\mathbb{R})$ to $U(1)$ mentioned before
\eqref{redEqns}.

The Killing horizons are determined by the zeroes of the Killing
norm. The norm squared is given by $k^\mu k^\nu g_{\mu\nu}=g_{tt}$,
and therefore by solving $g_{tt}=0$ the horizons are located at
\begin{equation}\label{eq:horizon}
r_h=L\,\Big[{E^2L^4\over 4}\pm\sqrt{\left({E^2L^4\over 4}\right)^2+M}\Big]~.
\end{equation}
For positive $M$, there is exactly one positive solution to
\eqref{eq:horizon}. If $M$ is negative two Killing horizons exist,
provided the inequality $E^2 > 4\sqrt{-M}/L^4$ holds. If the
inequality is saturated,
\begin{equation}
M_{\rm ext}=-\frac{L^8E^4}{16}~,
\end{equation}
then the Killing horizon becomes extremal and the value of the
dilaton \eqref{eq:gen2} on the extremal horizon, $r_h/L=L^4 E^2/4$,
coincides with the constant dilaton result \eqref{eq:dilaton}. This
is consistent with the universality of extremal black hole
spacetimes \cite{Kunduri:2007vf}.

The geometric properties of the solution \eqref{eq:lindil} are
developed further in \cite{Grumiller:2002nm,Grumiller:2006rc} and
references therein. The thermodynamic properties are a special case
of those discussed in \cite{Grumiller:2007ju}.\footnote{The solution
\eqref{eq:lindil} is the special case $U(X)=0$,
$V(X)=-\frac{4X}{L^2}+L^2E^2$ where the functions $U$, $V$ are
introduced in the definition of the action (1.1) of
\cite{Grumiller:2007ju} and $X=e^{-2\phi}$ is the dilaton field.}


\subsection{Asymptotic geometry and counterterms}
In order to compare the asymptotic geometry with our previous results
we introduce
\begin{equation}
e^{2\eta/L} = \frac{4r}{L} ~,
\label{eq:gen6}
\end{equation}
and write the solutions \eqref{eq:lindil} in the form
\begin{subequations}
\label{eq:geneom}
\begin{align}
g_{\mu\nu} dx^\mu dx^\nu & = d\eta^2 - \frac14 e^{4\eta/L} dt^2 + \dots~, \\
A_\mu dx^\mu &= \frac12 LE e^{2\eta/L} dt ~, \\
\phi &= -\frac{\eta}{L} ~.
 \label{eq:lalapetz}
\end{align}
\end{subequations}
At this order the solutions agree with the constant dilaton
background \eqref{Solution}, except the dilaton diverges linearly
with $\eta$ rather than approaching a constant. Therefore, the
solution \eqref{eq:lindil} may be called `linear dilaton solution'.

Asymptotically linear dilaton solutions have Fefferman-Graham like
expressions analogous to \eqref{eq:FG}, except that we must allow a
logarithmic modification in the expansion of the dilaton
\begin{subequations}
\label{eq:gen7}
\begin{align}
h_{tt} &= e^{4\eta/L}\,h_{tt}^{(0)} + h_{tt}^{(1)} + \dots~, \\
A_t &= e^{2\eta/L}\,A_t^{(0)} + A_t^{(1)} + \dots~, \\
\phi &= \eta\,\phi^{(\rm log)} + \phi^{(0)} + \dots~.
\end{align}
\end{subequations}
As for the full solution, the asymptotic geometry and field strength
of the linear dilaton solutions respect the asymptotic
$SL(2,\mathbb{R})$ symmetry, but the asymptotic dilaton respects only
the Killing vector $\partial_t$. The explicit linear dilaton solution
\eqref{eq:lindil} is obviously of the asymptotically linear dilaton
form form \eqref{eq:gen7}. Its boundary values are
\begin{equation}
h_{tt}^{(0)}=-\frac14\,,\qquad A_t^{(0)}= \frac12 LE\,,\qquad \phi^{(\rm log)} = -\frac 1L\,.
\label{eq:gen8}
\end{equation}

We want to set up a consistent boundary value problem and variational
principle, as in section \S\ref{sec:2}. There we wrote down the most
general local counter terms and determined their coefficients,
essentially by demanding the vanishing of the momenta
\eqref{eq:bndymom} at the boundary. A non-constant dilaton could give
rise to some additional local boundary terms, but all candidate terms
vanish too rapidly to affect the variational principle --- they are
irrelevant terms in the boundary theory. Since there are no new
counter terms and the coefficients of the existing ones are fixed by
considering constant dilaton configurations, it must be that the same
counter terms suffice also in the more general case.

We can verify this argument by explicit computation. For our choice
of counter terms, the momenta \eqref{Amomentum}, \eqref{phimomentum}
vanish no matter the dilaton profile. For consistency we have
verified that \eqref{hmomentum} also does not lead to new conditions.
We conclude that the logarithmic modification in the dilaton sector
inherent to \eqref{eq:gen8} does not destroy the consistency of the
full action \eqref{Action}. Moreover, the discussion of gauge
invariance in section \S\ref{sec:gaugeinv} also carries through. This
is so, because the new boundary term \eqref{eq:bm25} does not depend
on the dilaton field.

In summary, we find that the full action \eqref{Action} encompasses the
boundary value problems \eqref{eq:bc} and \eqref{eq:gen8}, has a
well-defined variational principle, and is extremized by the constant
dilaton solutions \eqref{Solution} as well as by the linear dilaton
solutions \eqref{eq:lindil}. Our discussion therefore exhausts all
solutions to the equations of motion \eqref{eq:eom}.

\section{Discussion}
\label{sec:7}
We conclude this paper with a few comments on questions that
are left for future work:

\begin{itemize}
\item {\bf Universal Central Charge:}
Our result  for the central charge can be written as \eqref{centcharge}
\begin{equation}\label{centcharg}
c =  \frac{3}{G_2}~.
\end{equation}
This form of the central charge does not depend on the detailed matter in the theory,
{\it i.e.} the Maxwell field and the charge of the solution under that field. This raises
the possibility that the central charge \eqref{centcharg} could be universal, {\it i.e.}
independent of the matter in the theory. It would therefore be interesting to study
more general theories and establish in which cases \eqref{centcharg} applies.

\item {\bf Mass Terms for Gauge Fields:} One of the subtleties we
    encountered in this paper was the presence of the mass term
    \eq{ I_{\rm new} \sim \int_{\dM}\bns m \,A^2~, }{eq:conc1}
    for the boundary gauge field. Related boundary terms are
    known from Chern-Simons theory in three dimensions
    \cite{{Gukov:2004id},{Kraus:2006nb}}, but apparently not in
    higher dimensions. It would be interesting to find situations
    where such boundary terms do appear in higher dimensions,
    after all. A challenge is that typical boundary conditions
    have the gauge field falling off so fast at infinity that
    these boundary terms are not relevant, but there may be
    settings with gauge fields that fall off more slowly.
\item
{\bf Unitarity:} Our computations have several unusual signs. The most prominent
one is that the overall constant in the action \eqref{eq:aa} must be negative
\begin{equation}
\alpha<0~.
\end{equation}
With this assignment the various terms in the action would appear
to have the ``wrong'' sign, raising concerns about the unitarity.
The sign we use is required to get positive central charge,
positive 3D Newton constant, positive energy of the 2D black
holes, and positive black hole entropy. This suggests that
$\alpha<0$ is in fact the physical sign. Nevertheless, a more
direct understanding of unitarity would be desirable.
\end{itemize}
We hope to return to these questions in future work.

\section*{Acknowledgements}
We thank A. Faraggi, R. Jackiw, P. Kraus, A. Maloney, S. Minwalla, A.
Sen and A. Strominger for discussions. We are grateful to J. L. Davis
and K. Hanaki for collaboration during the early phase of this paper.

Research at Perimeter Institute (PI) is supported by the Government
of Canada through Industry Canada and by the Province of Ontario
through the Ministry of Research \& Innovation. DG has been supported
by the project MC-OIF 021421 of the European Commission under the
Sixth EU Framework Programme for Research and Technological
Development (FP6). Research at the Massachusetts Institute of
Technology is supported in part by funds provided by the U.S.
Department of Energy (DoE) under the cooperative research agreement
DEFG02-05ER41360. The work of AC and FL is supported in part by DoE
under grant DE-FG02-95ER40899. DG acknowledges travel support by PI
and the kind hospitality at PI during the initial stage of this
paper. FL and RM would also like to thank the organizers of the
workshop ``Gravitational Thermodynamics and the Quantum Nature of
Space Time'' for hospitality during the early stages of this work.

\appendix

\section{Conventions and Notations}\label{app:not}

The 2D Newton's constant is determined by requiring that the
normalization of the gravitational action is given by
\begin{equation}\label{gravaction}
I=-{1\over 16\pi G_2}\int_\MM d^2x\, \sqrt{-g}\,R+\ldots,
\end{equation}
where the unusual minus sign comes from requiring positivity of
several physical quantities, as explained in the body of the paper.
Comparing \eqref{eq:bulkAction} with \eqref{gravaction} gives the
relation \eqref{eq:G2} for constant dilaton backgrounds.

For sake of compatibility with other literature we choose a somewhat
unusual normalization of the AdS radius $L$ so that in 2D
$R_{AdS}=-8/L^2$, and for electric field $E$ we use
$F_{\mu\nu}=2E\,\epsilon_{\mu\nu}$. For the same reason our gauge
field has inverse length dimension. As usual, the quantity
$F_{\mu\nu}=\partial_\mu A_\nu-\partial_\nu A_\mu$ is the field
strength for the gauge field $A_{\mu}$, its square is defined as
$F^2=F^{\mu\nu}F_{\mu\nu}=-8E^2$, and the dilaton field $\phi$ is
defined by its coupling to the Ricci scalar of the form
$e^{-2\phi}\,R$.

Minkowskian signature $-,+,\dots$ is used throughout this paper.
Curvature is defined such that the Ricci-scalar is negative for AdS.
The symbol $\MM$ denotes a 2D manifold with coordinates $x^\mu$,
whereas $\dM$ denotes its timelike boundary with coordinate $x^a$ and
induced metric $h_{ab}$. We denote the 2D epsilon-tensor by
\begin{equation}
\epsilon_{\mu\nu} = \sqrt{-g} \, \varepsilon_{\mu\nu}~,
\end{equation}
and fix the sign of the epsilon-symbol as $\varepsilon^{t\eta} =
-\varepsilon_{t\eta} =1$.

In our 2D study we use exclusively the Fefferman-Graham type
of coordinate system
\begin{equation}
ds^2=d\eta^2+g_{tt}\,dt^2~,
\end{equation}
in which the single component of the induced metric on $\dM$ is
given by $h_{tt} = g_{tt}$ with `determinant' $h=h_{tt}$. In the
same coordinate system the outward pointing unit vector normal to
$\dM$ is given by $n^{\mu} = \delta^{\mu}_{\eta}$, and the trace of
the extrinsic curvature is given by
\begin{equation}
K = {1\over2}h^{tt}\,\partial_{\eta}h_{tt}~.
\end{equation}

Our conventions in 3D are as follows. Again we use exclusively the
Fefferman-Graham type of coordinate system
\begin{equation}
ds^2=d\eta^2+\gamma_{ab}\,dx^a dx^b~,
\end{equation}
in which the induced metric on the boundary is given by the 2D
metric $\gamma_{ab}$. In the same coordinate system the extrinsic
curvature is given by
\begin{equation}
\Kthree_{ab}=\frac12 \partial_\eta \ga_{ab}~,
\end{equation}
with trace $\Kthree=\ga^{ab}\Kthree_{ab}$. The 3D AdS radius $\ell$
is normalized in a standard way, $\Rthree_{\rm AdS}=-6/\ell^2$.
Without loss of generality we assume that the AdS radii are
positive: $L,\ell>0$.

\section{Dictionary between 2D and 3D}\label{app:A}

We have derived in section \S\ref{sec:4.2} the relation
\eqref{diccscalar} between the normalization constants in 2D and 3D.
As a consistency check on our 3D interpretation of the 2D theory we
show in section \ref{app:A.1} that the boundary terms also reduce
correctly. Also, since the 2D Maxwell-dilaton theory is on-shell equivalent
to the KK-reduction of 3D gravity, the 3D solutions respecting the appropriate
isometry must agree with a 2D solution. We construct the
explicit map in section \ref{app:A.2}.

\subsection{Kaluza-Klein reduction: the boundary terms}\label{app:A.1}

Applying the KK-reduction \eqref{KKmetric} to a 3D metric in the
Fefferman-Graham form \eqref{FG} we can write
\begin{equation}\label{KKmetric2}
ds^2=e^{-2\psi}\ell^2(dz+\tilde{A}_t dt)^2+\tilde{h}_{tt}dt^2+ d\eta^2~.
\end{equation}
Here we identify $\tilde{h}_{tt}$ as the metric of the 1D boundary of
the 2D metric $\tilde{g}_{\mu\nu}dx^\mu dx^\mu$. Surfaces of
(infinite) constant $\eta$ define the boundary in both 3D and in 2D,
and so we can use $\eta$ as the radial coordinate in both cases. The
3D trace of extrinsic curvature becomes
\begin{equation}\label{3dK}
\Kthree=\tilde{K}-\partial_\eta \psi~,
\end{equation}
with $\tilde{K}$ the extrinsic curvature of the one dimensional
boundary  $\tilde{h}_{tt}$. The boundary term of the 3D theory in
(\ref{3dbulk}) therefore reduces to the boundary term
\begin{equation}\label{2dKKct}
\tilde{I}_{\rm boundary}={\ell\over4 G_3}\int
dt\sqrt{-\tilde{h}}\,e^{-\psi}\Big(\tilde{K}-{1\over\ell}\Big)~,
\end{equation}
of the 2D theory. The term proportional to the gradient in $\psi$
canceled an identical term arising when integrating the bulk term
\eqref{3dRicci} by parts.

In order to show that our 2D theory is equivalent on-shell to the
KK-reduction of the 3D theory we must match \eqref{2dKKct} with the
boundary term
\begin{equation}\label{2dct}
I_{\rm boundary}={\alpha\over\pi}\int dt\sqrt{-h}\,e^{-2\phi}\Big(K-{2\over
L}+{L\over2}e^{2\phi}A^aA_a\Big)
\end{equation}
determined directly in 2D. Evaluating (\ref{2dct}) on the asymptotic
AdS$_2$ backgrounds \eqref{Solution} we have
\begin{equation}\label{bndyKA}
K={2\over L}~,\qquad h^{ab}A_aA_b =-e^{-2\phi}{4\over L^2}~.
\end{equation}
and so
\begin{align}\label{onshellct}
I_{\rm boundary}&=-{\alpha\over\pi}\int dt \sqrt{-h}e^{-2\phi}{2\over L}\nonumber\\
&={\ell\over4 G_3}\int dt\sqrt{-\tilde{h}}e^{-\psi}{1\over\ell}~,
\end{align}
where in the last line we used our 3D-2D dictionary
(\ref{diccmectric}), (\ref{diccscalar}).

Asymptotically AdS$_2$ solutions of the theory defined by the bulk action (\ref{2dKKbulk}) have extrinsic
curvature $\tilde{K}={2\over\ell}$ and therefore the on-shell value
of the KK reduced boundary action (\ref{2dKKct}) exactly agrees with the on-shell value of the 2D boundary action (\ref{onshellct}), i.e.,
$\tilde{I}_{\rm boundary}=I_{\rm boundary}$. This is what we wanted to show.

\subsection{Asymptotically AdS solutions}\label{app:A.2}

Our starting point is the general 3D solution (\ref{FGAdS3}). For
compactifications along $z=x^+$ we consider $\FGA=\rm constant$, and
rewrite the solution in the form (\ref{KKmetric2}) as
\begin{eqnarray}
ds^2&=&\Big({\formerg\over\ell}\Big)\ell^2\Big[dx^+
+{1\over2\ell}e^{2\eta/\ell}\sqrt{\ell\over
\formerg}\big(1+{16\over\ell^2}\formerg\bFGA(t)e^{-4\eta/\ell}\big)dt\Big]^2\nonumber\\
&~&~~~~
-{1\over4}e^{4\eta/\ell}\big(1-{16\over\ell^2}\formerg\bFGA(t)e^{-4\eta/\ell}\big)^2dt^2
+d\eta^2~,\label{FGnew}
\end{eqnarray}
with
\begin{equation}\label{tcoordef}
t={\ell\over4}\sqrt{\ell\over\formerg}\,x^-~.
\end{equation}
Comparing with the Ansatz \eqref{KKmetric} we read off the 2D metric
\begin{equation}\label{FGmetric}
d\tilde{s}^2=\tilde{g}_{\mu\nu}dx^\mu
dx^\nu=-{1\over4}e^{4\eta/\ell}\Big(1-{16\over\ell^2}\formerg\bFGA(t)e^{-4\eta/\ell}\Big)^2dt^2
+d\eta^2~,
\end{equation}
and the matter fields
\begin{subequations}\label{FG2d}
\begin{align}
\tilde{A}&={1\over2\ell}e^{2\eta/\ell}e^{\psi}\Big(1+{16\over\ell^2}\formerg\bFGA(t)e^{-4\eta/\ell}\Big)dt~,\\
e^{-2\psi}&={\formerg\over \ell}~. \label{psidiv}
\end{align}
\end{subequations}
The solution \eqref{FGmetric}-\eqref{FG2d} should be equivalent to
the asymptotically AdS$_2$ solutions \eqref{Solution} found directly
in 2D. After the coordinate transformation $(\eta,\,t)\to {1\over
a}\,(\eta,\,t)$ in \eqref{Solution} this expectation is correct,
and we use the dictionary (\ref{dicc}), (\ref{diccscalar}) to
find the relations $h_0(t)=1$, $a(t)=0$ and
\begin{subequations}\label{h1}
\begin{align}
h_1&=-{16\over\ell^2}\formerg\bFGA(t)~,\label{ctwodic} \\
\alpha &= -{\pi\ell\over 8G_3} e^{2\phi} \sqrt{\frac{\formerg}{\ell}} ~, \label{eq:anotherlabel}
\end{align}
\end{subequations}
between the parameters of the solutions.

\bibliographystyle{fullsort}
\bibliography{review}
\end{document}